\documentclass[useAMS,usenatbib]{mn2e}

\usepackage{epsfig, natbib, color}

\newcommand{\hii}{H~\textsc{ii}}
\newcommand{\ft}{f_{\rm trap}}
\newcommand{\calE}{\mathcal{E}}
\newcommand{\msun}{M_{\odot}}
\newcommand{\epsff}{\epsilon_{\rm ff}}
\newcommand{\epsfftwo}{\epsilon_{\rm ff,-2}}

\newcommand{\lsun}{L_\odot}
\newcommand{\ifm}[1]{\relax\ifmmode#1\else$\mathsurround=0pt #1$\fi}
\newcommand{\kms}{\ifmmode\,{\rm km}\,{\rm s}^{-1}\else km$\,$s$^{-1}$\fi}

\newcommand{\kpc}{\,{\rm kpc}}

\newcommand{\ltsima}{$\; \buildrel < \over \sim \;$}
\newcommand{\lsim}{\lower.5ex\hbox{\ltsima}}
\newcommand{\gtsima}{$\; \buildrel > \over \sim \;$}
\newcommand{\gsim}{\lower.5ex\hbox{\gtsima}}

\newcommand{\equ}[1]{eq.~(\ref{eq:#1})}

\newcommand{\se}[1]{\S\ref{sec:#1}}

\newcommand{\be}{\begin{equation}}
\newcommand{\ee}{\end{equation}}
\newcommand{\bea}{\begin{eqnarray}}
\newcommand{\eea}{\end{eqnarray}}

\def\sy{\,M_\odot\, {\rm yr}^{-1}}

\def\Mv{M_{\rm v}}

\def\Md{M_{\rm d}}
\def\Rd{R_{\rm d}}
\def\Sigmac{\Sigma_{\rm crit}}
\def\Sigmay{\Sigma_{\rm crit,y}}

\title[Giant Clump Survival]{Survival of Star-Forming Giant Clumps in High-Redshift Galaxies}

\author[Mark R. Krumholz and Avishai Dekel]{Mark R. Krumholz$^1$\thanks{Email: krumholz@ucolick.org (MRK); dekel@phys.huji.ac.il (AD)} and Avishai Dekel$^2$\footnotemark[1]\\
$^1$Department of Astronomy \& Astrophysics, University of California, Santa
Cruz, CA 95060, USA\\
$^2$Racah Institute of Physics, The Hebrew University, Jerusalem 91904, Israel}

\begin{document} 

\date{Accepted 12 March 2010}

\pagerange{\pageref{firstpage}--\pageref{lastpage}} \pubyear{2010}

\maketitle

\label{firstpage}

\begin{abstract}
We investigate the effects of radiation pressure from stars
on the survival of the star-forming giant clumps in high-redshift massive 
disc galaxies, during the most active phase of galaxy formation.
The clumps, typically of mass $\sim 10^8-10^9\msun$ and radius $\sim 0.5-1\kpc$,
are formed in the turbulent gas-rich discs by violent gravitational instability
and then migrate into a central bulge in $\sim 10$ dynamical times.
We show that the survival or disruption of these clumps under the 
influence of stellar feedback depends critically on the rate at which 
they form stars. If they convert a few percent of their gas mass to 
stars per free-fall time, as observed for all local 
star-forming systems and implied by the Kennicutt-Schmidt law, they cannot 
be disrupted. Only if clumps convert most of their mass to stars in a few free-fall times
can feedback 
produce significant gas expulsion. We consider whether such rapid star 
formation is likely in high-redshift giant clumps.
\end{abstract}



\section{Introduction}
\label{sec:intro}

A significant fraction of the massive galaxies, $\sim 10^{11}\msun$ 
in baryons, during the period from $z=1.5-3$ when star formation
is at its peak and most stellar mass is assembled \citep{hopkins06a,magnelli09a},
form stars at high star formation rates (SFR) of $\sim 100 \sy$.
Many of these are turbulent, gas-rich, extended rotating discs 
in which much of the star formation takes place in a few {\it giant clumps}
\citep{Cowie95a, van-den-bergh96a, elmegreen04b,
elmegreen05a,elmegreen07a}.
Based on this morphology, they were first termed
``chain" or ``clump-cluster" galaxies.
In a typical galaxy of this type, $10-40\%$ of the UV rest-frame
light is emitted from a few clumps of 
characteristic size $\sim 1\kpc$ that form stars at tens of $\sy$ each 
\citep{elmegreen04a, elmegreen05a, forster-schreiber06a,genzel08a}. 
These clumps are much more massive than the star-forming
complexes in local galaxies.
The star formation in these clumps, and their survival subject to stellar
feedback, are central to our understanding of galaxy formation.

Kinematically, most of the galaxies that host the clumps 
are thick rotating discs, with high velocity dispersions of 
$\sigma = 20 - 80 \kms$ 
(one dimensional), compared to $\sigma \simeq 10\kms$ is present-day discs;
they have rotation to dispersion ratios of $V/\sigma \sim 1-7$ 
\citep{Cresci09a}. Estimates of the total gas fraction in star-forming galaxies, based on 
CO measurements, range from 0.2 to 0.8, with an average of 
$\sim 0.4-0.6$ \citep{tacconi08a, daddi08a, daddi09a}, systematically 
higher than the typical gas fraction
in today's
discs. These properties are generally incompatible with these systems 
being ongoing major mergers or remnants of such mergers 
\citep{Shapiro08a,Bournaud08a,dekel09b,Bournaud09a,dekel09a},
though there may be counter examples \citep{Robertson08a}.

Instead, giant clumps form through a scenario, 
summarized by \citet{dekel09a}, 
in which massive galaxies at $z\sim 2-3$ are fed by a few narrow and 
partly clumpy streams of cold gas ($\sim 10^4$K) that flow along 
the dark matter filaments of the cosmic web
\citep[e.g.][]{hahn07a}
and penetrate deep into the centres of the massive dark matter haloes
of $\sim 10^{12}\msun$ 
\citep{birnboim03a,keres05a,dekel06a,ocvirk08a,dekel09b}. 
Indeed, the existence of these streams is an {\it inevitable} prediction
of the standard $\Lambda$CDM cosmology.
They may produce the structures we observe as
Lyman-alpha blobs \citep{furlanetto05a, goerdt09a, dijkstra09a}.
The angular momentum they carry leads the accreted material 
to form a disc of radius $\Rd \sim 10\kpc$. 
The continuous intense input of gas at the level of $\sim 100 \sy$
maintains a high gas surface density $\Sigma$,
which drives a violent gravitational instability 
with a Toomre Q parameter
below unity, $Q \simeq \sigma \Omega /(\pi G \Sigma) <1$, 
where $\sigma$ is
the one-dimensional
velocity dispersion and $\Omega$ is the angular velocity
associated with the potential well \citep{toomre64}.
The disc instability is self-regulated at $Q \lsim 1$ by the gravitational
interactions in the perturbed disc, 
which keep the disc thick and with a high velocity dispersion.
The disc forms strong, transient spiral features that
fragment to produce
5-10 bound clumps,
that together comprise $\sim 20\%$ of the disc mass.
The largest clumps have characteristic radii
$R\simeq 7G\Sigma/\Omega^2\sim 1$ kpc, 
and characteristic masses of 
a few percent of the disc mass, $M\sim 10^9$ $\msun$.
A spectrum of smaller clumps with somewhat lower masses
forms as well.
The clumps' large 
masses cause them to migrate to the centre of the disc on a
short time scale of $\sim 10$ disc crossing times, 
where they merge into a central bulge 
\citep{noguchi99a,immeli04a,immeli04b,bournaud07a,ceverino09b,agertz09a}. 
This violent instability phase
can last for more than a gigayear, 
during which the mass flow from the disc to the bulge is 
replenished by fresh accretion, keeping
the mass within the disc 
radius divided quite evenly between disc, bulge and dark matter components.

While this scenario of clump formation and migration to build up bulges 
is appealing, it relies on the ability of clumps to survive for 
$\sim 10$ disc dynamical times, i.e.\ a few hundred million years, 
while the gas in them turns into stars on a comparable timescale 
\citep{genzel08a,dekel09a,ceverino09b}. However, at the SFR of a few 
tens of solar masses per year in the clumps, it is possible that they 
might be disrupted by stellar feedback on considerably shorter timescales. 
\citet{murray09a} argue for exactly this scenario. In their models, 
clumps disrupt after $\sim 1$ dynamical time, during which they turn only 
$\sim 30\%$ of their mass into stars. In this picture clumps would not 
survive long enough to migrate, and bulges would instead need to be built 
up by mergers. While this scenario seems difficult to reconcile with the 
estimated ages of a few hundred Myr for the oldest stellar populations in 
some clumps \citep{elmegreen09b, forster-schreiber09a}, there is 
sufficient uncertainty in both the observational estimates of clump ages 
and the theoretical modeling of clump evolution and disruption to merit 
a re-investigation of the problem, which we provide in this paper.

The outline of the paper is as follows. 
In \se{survival} we derive the expected gas ejection fraction as a 
function of SFR efficiency. 
In \se{clumpsfr} we address the observational estimates of the SFR 
efficiency. 
In \se{discussion} we discuss some of the issues raised by our results 
and compare to previous work. 
We summarize our conclusions in \se{summary}.

\section{Radiative Feedback and Clump Survival}
\label{sec:survival}

Consider a uniform-density giant gas clump of mass $M$, 
radius $R$, and surface density $\Sigma = M/(\pi R^2)$. 
It forms stars at a rate $\dot{M}_*$, and the stars formed within it have 
a combined luminosity $L$. (We defer discussing
the effect of sub-clumping within giant clumps to \S~\ref{sec:subclumps}, 
since it does not change the qualitative result.)
Characteristic numbers to keep in mind
for the largest, best-observed clumps, found in galaxies
with baryonic masses $\sim 10^{11}$ $\msun$,
are 
$M \simeq 10^9$ $\msun$, $R \simeq 1$ kpc, and $\Sigma \simeq 0.1$ g cm$^{-1}$.
Clumps in the $\sim 10^{10}$ $\msun$ galaxies,
which are more common, have masses $\sim 10^8$ $\msun$ and
sizes that are at or below the resolution limit of present observations. We
will assume that they have surface densities comparable to those their
larger cousins; they cannot be much smaller, since the mean column densities
of the galactic disks as a whole is $\Sigma\sim 0.05$ g cm$^{-2}$.
We wish to evaluate the fraction $e$
of clump mass that will be ejected
by stellar feedback and the fraction $\calE=1-e$ that is 
transformed into stars, because this is the critical parameter 
that determines
whether the clump will form a bound stellar system. Both
N-body simulations and analytic models
\citep[e.g.][]{hills80, kroupa01a, kroupa02a, baumgardt07a} 
indicate that if $\calE \ga 0.5$, then
most of the stellar mass will remain bound, while if $\calE \la 0.3$
then no bound stellar system will be left. 
Small portions of the clump where $\calE$ was
locally higher may form  bound clusters, but these will be 
orders of magnitude smaller than the initial clump.

Several common feedback mechanisms are not important for giant clumps in the relevant mass range. First, supernova feedback is unlikely to be effective in ejecting mass from these giant clumps, due to cooling and leakage of hot gas \citep{dekel09a,krumholz09d,murray09a}. Second, the pressure of warm ($\sim 10^4$ K) ionized gas is ineffective because the escape velocity from the clump is larger than the gas sound speed of $\sim 10$ km s$^{-1}$. Third, protostellar outflows are unable to eject mass  because they do not provide enough momentum \citep{fall10a}. Instead, the dominant feedback mechanism is likely to be radiation pressure from newly-formed stars, which creates a radiation-dominated \hii\ region. The expansion of such a region follows a similarity solution \citep{krumholz09d}, and \citet{fall10a} use this solution to show that all the remaining gas will be ejected once the fraction of gas mass transformed into stars reaches a value
\begin{equation}
\label{efflum}
\calE = \frac{M_*}{M} = \frac{\Sigma}{\Sigma + \Sigma_{\rm crit}},
\end{equation}
where
\begin{equation}
\Sigmac = \frac{5 \eta \ft \left\langle L/M_*\right\rangle}
{\pi \alpha_{\rm crit} G c},
\end{equation}
$\eta$ is a constant of order unity that depends on the density distribution 
within the clump, $\left\langle L/M_*\right\rangle$ is the light to mass ratio
of the stellar population, $\ft$ represents the factor by which the radiation 
force is enhanced by trapping of re-radiated infrared light within the 
expanding shell, and $\alpha_{\rm crit}$ is a parameter of order unity that 
describes the critical velocity required to eject mass from the cloud. 
\citet{fall10a} assume fiducial values of $\eta=2/3$ and 
$\ft=\alpha_{\rm crit} = 2$ for the local star-forming regions, 
and we adopt the same values here since there is no obvious reason for them
to be systematically different in the case of giant clumps at redshift 2.
We refer to $\calE$ as the final star fraction.\footnote{Note that 
this is sometimes referred to as star formation efficiency as well, but 
we avoid the term efficiency because it does not have a standard meaning, 
and different authors use it for different concepts.}

To determine the light to mass ratio, we must deal with the complication 
that the characteristic crossing time of a giant clump is rather long,
$t_{\rm cr} = R/\sqrt{GM/R} = 15 M_9^{-1/2} R_1^{3/2}$ Myr, 
where $M_9 = M/(10^9\,\msun)$ and $R_1=R/(1\mbox{ kpc})$. 
Depending on the exact values of $M$ and $R$, this can be either greater 
than or less than the main-sequence lifetime of massive stars. Thus we can 
neither assume a single burst, so that all stars are coeval, nor continuous star formation, so that the population is in equilibrium between new stars forming and old ones evolving off the main sequence. 
However, we can treat these scenarios as two limiting cases, which must bracket any real stellar population. \citet{krumholz07e} point out that in stellar populations younger than
3 Myr, where no stars have left the main sequence yet, the light-to-mass 
ratio has a constant value, $\Psi$, for which we adopt 
$\Psi\approx 2200$ erg s$^{-1}$ g$^{-1}$ \citep{fall10a}. 
Once a stellar population is old enough to reach statistical equilibrium 
between star formation and star death, it instead has a nearly constant luminosity-to-star-formation 
rate ratio, $\Phi$, which we take to be 
$\Phi \approx 6.1\times 10^{17}$ erg g$^{-1}$ \citep{krumholz07e}. 
For any realistic stellar population whose light is dominated by young, massive stars as opposed to 
old ones, the luminosity is roughly equal to the smaller of these two limits, 
and for simplicity we simply take the light-to-mass ratio to be
\begin{equation}
\left\langle \frac{L}{M_*}\right\rangle = 
\min\left(\Psi, \Phi \frac{\dot{M}_*}{M_*} \right) .
\end{equation}
We refer to the first case as the ``young stars" limit and the second as the ``old stars" limit, since they represent the opposite extremes of stellar populations that have undergone no evolution and populations that are old enough to have reached equilibrium between star formation and stellar evolution.

Substituting these two light-to-mass estimates into equation (\ref{efflum})
gives
\begin{equation}
\label{eff1}
\calE = \max \left[ \left(1 + \frac{\Sigmay}{\Sigma}\right)^{-1}, 1 - \frac{\Sigmay}{\Sigma} \left(\frac{\Phi}{\Psi t_{\rm dep}}\right)\right],
\end{equation}
where 
\begin{equation}
\label{sigmayoung}
\Sigmay = \frac{5\eta \ft \Psi}{\pi \alpha_{\rm crit} G c} \approx 1.2\mbox{ g cm}^{-2}
\end{equation}
is the critical density evaluated in the young-stars limit 
and $t_{\rm dep}=M/\dot{M}_*$
is the depletion time, i.e.\ the time that would be required to convert all of 
the gas into stars. The numerical evaluation in equation (\ref{sigmayoung}) 
is for the fiducial parameters of \citet{fall10a}. Note that this expression 
is a maximum rather than a minimum because $\calE$ is a decreasing function 
of $\langle L/M_*\rangle$.

\begin{figure}
\includegraphics{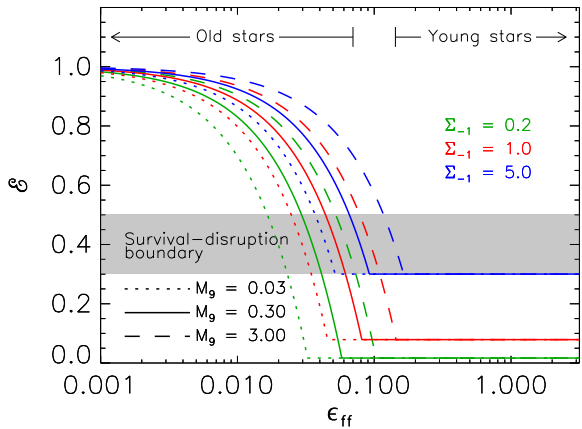}
\caption{
\label{fig:starfrac}
Final stellar fraction $\calE$ as a function of star formation rate efficiency 
$\epsff$, computed from Equation (\ref{eff2a}). We show results for 
giant clump surface densities $\Sigma_{-1} = 
0.2$, $1.0$, and $5.0$
 (green, red, and blue lines, as indicated), and for giant clump
masses $M_9 =
0.03$, $0.3$, and $3.0$
(dotted, solid, and dashed lines, as indicated). 
The arrows schematically indicate the region
around $\epsff = 0.1$ where we change from
the old stars limit (low $\epsff$) to the young stars limit (high $\epsff$).
The gray region from $\calE=0.3 - 0.5$ indicates the rough boundary
between stellar fractions $\calE \la 0.3$, for which no bound stellar
system will remain, and $\calE \ga 0.5$, for which a majority of the
stellar mass will remain bound.
}
\end{figure}

Following \citet{krumholz05c}, 
we define the dimensionless star-formation rate efficiency as the ratio 
between the free-fall time and the depletion time\footnote{Note that the 
rate efficiency $\epsff$ that we
have defined here is distinct both from the star formation rate, which
has units of $\msun$ yr$^{-1}$ and is not normalized by the
ratio of mass over free-fall time, and the final stellar mass fraction $\calE$,
which is dimensionless but does not carry any information about 
the star formation rate.}, namely 
\be
\epsff = \frac{\dot{M}_*}{M/t_{\rm ff} }.
\ee 
\citet{krumholz07e} show that $\epsff \simeq 0.01$ 
across a very broad range of densities, size scales, and environments.
Here we define the free-fall time as $t_{\rm ff} = \sqrt{3\pi/(32 G \rho)}$, 
where $\rho$ is the gas density. We discuss in \S~\ref{sec:clumpsfr} below
whether this value of $\epsff$ applies in high-$z$ giant clumps. 
If we adopt it for now and make this substitution in equation (\ref{eff1}), 
we find that
\begin{eqnarray}
\calE & = & \max \left[ \left(1 + \frac{\Sigmay}{\Sigma}\right)^{-1}, 1 - \frac{\epsff \sqrt{8G}\Sigmay \Phi}{(\pi \Sigma M)^{1/4}\Psi}\right]
\label{eff2a}
\\
& \simeq & \max\left[ \left(1 + \frac{12}{\Sigma_{-1}}\right)^{-1}, 1 - \frac{0.086}{(\Sigma_{-1} M_9)^{1/4}} \epsfftwo \right],
\label{eff2}
\end{eqnarray}
where $\Sigma_{-1} = \Sigma/(0.1\mbox{ g cm}^{-2})$, $\epsfftwo=\epsff/100$, 
and the numerical evaluation uses the fiducial parameters from \citet{fall10a}. 
We use Equation (\ref{eff2}) to plot $\calE$ as a function of $\epsff$
in Figure \ref{fig:starfrac}. 

What does this result imply for the survival of high-$z$ giant clumps? 
We note that the second term in brackets in equation (\ref{eff2}), 
corresponding to the case of a stellar population older than $\sim 3$ Myr, 
is generally the one that applies for giant clumps. This reflects the
fact that the crossing time for our fiducial values of the 
parameters is 15 Myr, which is significantly larger than 3 Myr. 
The young stellar population limit applies only if 
the star formation rate efficiency is much higher
than is observed in any star-forming systems anywhere in
the local universe, $\epsfftwo \ga 10$,
or if the star-forming systems are significantly less massive or much more 
dense.
This makes high-$z$ giant clumps very different from Galactic star clusters 
or even super-star clusters, such as those found in local starburst 
galaxies, e.g.\ the Antennae or M82.
These have lower masses and higher surface densities, 
with crossing times $\sim 0.1$ Myr \citep[e.g.][]{mccrady07a}, 
placing them firmly in the young-star limit. 
Indeed, the first term in the brackets is identical to 
that derived by \citet{fall10a} for local star-forming clumps.\footnote{To 
get numerical agreement in the coefficient of $\Sigma_{-1}$, we must adjust 
$k$ by a factor of $\sqrt{0.4}$ to account for measuring the escape 
velocity at the surface instead of at the half-mass radius.}

We conclude that unless $\epsfftwo \gg 1$ for giant clumps, 
as opposed to $\sim 1$ for the local star-forming systems, 
star-forming clumps with masses $\sim 10^8-10^9$ $\msun$ and surface densities 
$\sim 0.1$ g cm$^{-2}$ cannot be disrupted by radiation pressure
--- so they end up converting most of their mass to stars.
Thus,
the {\it expulsion fraction\,} relevant for high-$z$ giant clumps, as derived
in the old-star limit, is 
\be 
e = 1 - \calE = 0.086 \Sigma_{-1}^{-1/4} M_9^{-1/4} \epsfftwo \, .
\label{eq:expulsion}
\ee

Since $\calE \ga 0.5$, we expect the resulting stellar
systems to remain gravitationally bound. Clumps 
with $\Sigma_{-1}=1$ have
$\calE < 0.5$ and suffer significant disruption only if
their masses are below $\sim 10^6$ $\msun$, and they
reach $\calE < 0.3$ and undergo complete disruption only
at masses $\la 2\times 10^5$ $\msun$. Thus the
observed $\sim 10^8-10^9$ $\msun$ giant clumps 
in high$-z$ galaxies should survive disruption unless 
$\epsfftwo \gg 1$. Even if we are maximally conservative and
assume that the smaller clumps have $\Sigma_{-1} = 0.5$,
i.e.\ that their surface densities do not exceed the mean
surface densities of their host galaxies, our estimated maximum mass
for disruption only increases by a factor of 2.

\section{The Star Formation Rate Efficiency in High-z Clumps}
\label{sec:clumpsfr}

Equation (\ref{eq:expulsion}) shows that the most important parameter in 
determining the survival of giant clumps is how quickly, normalized to 
their free-fall times, they turn themselves into stars. Only if they do 
so with a very high rate efficiency,
$\epsfftwo \ga 10$, do we expect significant gas expulsion.
In the local universe, one measures $\epsff$ by determining the star 
formation rate of an object or a population of objects, and comparing 
this to the objects' gas mass divided by the free-fall time computed 
for their density. This procedure was first applied by \citet{zuckerman74} 
to the population of 
giant molecular clouds in the Milky Way, and has subsequently been 
extended to other objects by \citet{krumholz07e} and \citet{evans09a}. 
These measurements give $\epsfftwo \sim 1$ over a very wide range of 
star-forming environments, from small star clusters in the Milky Way to 
entire starburst galaxies. The results are subject to considerable 
uncertainty, but values $\epsfftwo \ga 10$ are strongly excluded by the 
data. However, we lack comparable data for star formation at high
redshifts. 
In this section, we therefore turn to the question of the likely value 
of $\epsff$ in high-$z$ giant clumps.

\subsection{Estimates of $\epsff$ from Observations of Giant Clumps}

Unfortunately, we cannot easily apply the direct measurement procedure 
used to determine $\epsff$ in the local universe to the 
high-$z$ clumps, because, while we can evaluate star-formation rates using 
H$\alpha$ luminosities, we are limited in our ability to measure the 
corresponding gas properties.

For example, \citet{elmegreen09c}\ use stellar population synthesis to estimate 
masses and ages for the stellar populations seen in giant clumps in high-$z$ 
galaxies, and they then compare the ages $\tau$ to the clump dynamical times, 
defined as $t_{\rm dyn} \equiv (G\rho)^{-1/2}\approx 0.5 t_{\rm ff}$, 
where $\rho$ is taken to be the {\it stellar} mass density. 
They find typical values $\tau/t_{\rm dyn}\sim 1-10$, corresponding to 
$\tau/t_{\rm ff}\sim 2-20$, with a factor of $\sim 10$ scatter. 
It is tempting to identify $\tau$ with $t_{\rm dep}$ and
simply estimate $\epsff \sim t_{\rm ff}/\tau \sim 0.05-0.5$, but this is likely
to be a significant overestimate. Based on dynamical mass estimates, 
\citet{genzel08a} estimate that gas comprises 10-30\% 
of the total mass within the disc radius. Estimates based on direct
CO measurements indicate gas fractions of $45-60\%$ at
$z \sim 2$ 
\citep{daddi09a,tacconi10a}.
It is likely to be an even larger fraction of the mass within the dense, 
rapidly star-forming giant clumps, which are self-gravitating and lack
a dark matter component.  If the stars measured by 
\citeauthor{elmegreen09c}\ comprise only a small fraction $f_*$ 
of the total mass 
in clump, then $\epsff$ will be reduced by a factor of roughly $f_*^{1.5}$ 
relative to the previous estimate -- one power of $f_*$ to account for the 
gas that has not yet formed stars, 
and another factor of $f_*^{0.5}$ 
because the free-fall time will be shorter than the value 
\citeauthor{elmegreen09c}\ estimate based on the stars alone. 
To give a sense of the possible magnitude of the error, note that 
\citet{murray09a} estimate $f_* = 0.2$ for a giant clump in BX 482 
(and we argue in Section \ref{murraymass} that $f_*$ is probably even smaller), 
and this value of $f_*$ would be sufficient to lower the estimate of 
$\epsff$ by a factor of 10, to 
$5\times 10^{-3} - 5\times 10^{-2}$. A secondary worry is that,
as \citeauthor{elmegreen09c}\ point out, 
since the clumps are selected using rest-frame blue light there is a strong 
bias against selecting older, redder clumps,
causing an underestimate of $\tau$. 
In general, $\tau$ is expected to be an underestimate of $t_{\rm dep}$
because the former refers only to the stars that have already formed.

Given this problem, many observers have attempted to estimate
gas masses via the 
``inverse" \citet{kennicutt98a} law, namely by 
measuring the SFR surface density and then assuming that the gas surface
density has the value required for the object to obey the Kennicutt law
\citep[e.g.][]{genzel06a, genzel08a}. 
Unfortunately this procedure does not yield an independent estimate of 
$\epsff$. The \citeauthor{kennicutt98a} relation is 
\begin{equation}
\label{kennicutt}
\dot{\Sigma}_* = A \Sigma_{-1}^{1.4},
\end{equation}
with $A\simeq 1.4$ $\msun$ yr$^{-1}$ kpc$^{-2}$. Thus if the scale height of 
the galaxy is $h$, then the midplane gas density is $\rho=\Sigma/h$, 
the free-fall time is $t_{\rm ff} = \sqrt{3\pi h/(32 G \Sigma)}$, and we have
\begin{equation}
\label{kennicutteps}
\epsfftwo = 100 \frac{\dot{\Sigma}_*}{\Sigma/t_{\rm ff}} 
= 3.4 \, A_{1.4} \,  h_0^{0.5} \Sigma_{-1}^{-0.1},
\end{equation}
where $h_0 = h/\mbox{kpc}$ and $A_{1.4}= A/(1.4$ $\msun$ yr$^{-1}$ kpc$^{-2})$. 
The true value of $\epsff$ is almost certainly a bit smaller than this, 
since star-forming clouds have densities higher than the mean midplane density,
and thus smaller free-fall times. Nonetheless, this calculation illustrates
a crucial point: to the extent that galactic scale heights do not have a very 
large range of variation (and the dependence is only to the 0.5 power), 
the statement that $\epsfftwo \sim 1$ is roughly equivalent to the Kennicutt 
law. (See \citealt{krumholz05c} and \citealt{schaye08a} for a more detailed 
discussion of the relationship between volumetric and areal star formation 
laws.) Thus, no measurement of the gas surface density that assumes the 
Kennicutt law {\it a priori} can produce a value of $\epsfftwo$ significantly 
different from this. Any measurement of $\epsfftwo$ that did 
yield a significantly larger value would necessarily place the galaxy well 
off the Kennicutt relation.

\subsection{Estimates of $\epsff$ from Observations of Clump Host Galaxies}

Given the difficulties of estimating $\epsff$ in giant clumps directly, 
we instead turn to indirect inferences,
based on the more robust measurements of the overall disc properties.
We first note that the observed correlation between total gas mass
and total star-formation 
rate in high-$z$ galaxies \citep[e.g.][]{carilli05, greve05, gao07a} 
is consistent with these galaxies having the same value of $\epsff$ as local
star-forming systems \citep{krumholz07g, narayanan08a, narayanan08b}. 
\citet{bothwell09a} have claimed to detect a deviation 
from a universal star formation law in spatially resolved observations of 
three $z\sim 2$ systems. However, 
\citeauthor{bothwell09a} obtain this result only because they
choose a non-standard 
conversion factor between CO luminosity and mass, which leads 
them to conclude that the gas fraction in these systems is 
only $\sim 5-10\%$, much lower than in typical star forming 
galaxies at $z\sim 2$, and on the low side even for local star-forming galaxies.
\citet{tacconi10a}, using a standard conversion factor,
conclude instead that these galaxies have molecular gas
fractions of $35-45\%$. This is consistent with 
the star formation law and the value of $\epsff$ at $z\sim 2$ being the same 
as in the local universe.

We can also approach the problem more theoretically. The high-$z$
clumps discs with which we are concerned have
SFR $\dot{M}_* \sim 100\sy$, baryonic mass $\Md \sim 10^{11}\msun$,
and disc radius $\Rd \sim 10\kpc$.
We first express the relevant quantities in Equation (\ref{eq:expulsion})
as a function of the clump mass $M$, 
surface density $\Sigma$, and SFR $\dot{M_*}$. 
The free-fall time (which is $\sqrt{5}/2$ times the crossing time $R/V$),
expressed in $10^6$ yr, is 
\be
t_{{\rm ff},6} \equiv \sqrt{\frac {3\pi}{32G\rho}}
= 16.7 M_9^{-1/2} R_1^{3/2} 
= 12.3 M_9^{1/4} \Sigma_{-1}^{-3/4} \, .
\label{eq:tff}
\ee
Then
\be
\epsfftwo \equiv \frac{\dot{M}_*}{M/t_{\rm ff}}
=12.3 \dot{M}_{*10} \Sigma_{-1}^{-3/4} M_9^{-3/4} \, ,
\label{eq:eps}
\ee
where $\dot{M}_{*10}$ is the SFR in a clump in $10\sy$.

Now let us express the clump quantities in terms of the disc mass, 
surface density, and total SFR.
The disc surface density is
\be
\Sigma_{d,-1}= 0.663 M_{d,11} R_{d,10}^{-2} \, ,
\ee
where $R_{d,10}$ is the disc radius measured in units of 10 kpc.
The virialized clumps can be assumed to have collapsed by at least a factor 2 
in radius, so the clump surface density is $s\equiv 4 s_{4}$ times 
larger than the disc's, or 
\be
\Sigma_{-1} = 2.65 s_4 M_{d,11} R_{d,10}^{-2} \, ,
\label{eq:Sigma}
\ee
with $s_4 \geq 1$.

Assume that a fraction $\alpha$ ($\simeq 0.2$) of the disc mass is in $N$
($\sim 10$) identical clumps and a fraction $\beta$ ($\simeq 0.5$) of the 
SFR is in the clumps \citep{dekel09a,ceverino09b}.
Then
\be 
\dot{M}_{*,10} = 0.5 \beta_{0.5} N_{10}^{-1} \dot{M}_{d*,100} \, ,
\ee
and
\be
M_9=2 \alpha_{0.2} N_{10}^{-1} M_{d,11}\, .
\ee
Substituting the last three expressions in \equ{eps} we obtain in terms of the
disc quantities
\be
\epsfftwo = 1.75 s_4^{-3/4} \beta_{0.5} \alpha_{0.2}^{-3/4} 
N_{10}^{-1/4} \dot{M}_{d*,100} M_{d,11}^{-3/2} R_{d,10}^{3/2} \, .
\ee   
Note the very weak dependence on $N$.
We see that the most straightforward observational estimates for the massive
clumpy discs at $z \sim 2$ \citep[e.g.\ the BzK galaxies,][]{genzel08a} yield
$\epsfftwo \simeq 1.75$ and $e \sim 0.1$,
i.e.\ no significant expulsion.

This estimate is based on the observed properties of galaxies over 
a relatively narrow 
range of galaxy masses and redshifts, but using
simple theoretical arguments we can deduce how the results are likely 
to scale to other galaxies for which we currently lack direct
observations. 
For a self-gravitating disc, the circular velocity is roughly 
$V_{d,200}^2 \simeq M_{d,11}/R_{d,10}$, 
so $\epsfftwo \propto \dot{M}_{d*} V_{d}^{-3}$, namely
\be
\epsfftwo \propto \frac{\dot{M}_{d*}}{M_{\rm vir}} \, .
\ee
A constant $\epsfftwo$ ($\sim 1$) is consistent with 
the SFR being a constant fraction of the baryon accretion rate,
because the latter is roughly proportional to halo mass 
\citep{neistein06a,birnboim07a}:
\be
\dot{M}_{d*} \sim \dot{M} \propto M_{\rm vir}. 
\ee
The fact that the SFR follows the accretion rate is 
a natural result of the fact that the SFR is proportional to the mass
of the available gas \citep{bouche09a,dutton09b}, and is
consistent with the finding from simulations when compared to observed
SFR \citep{dekel09b}. 
We learn that $e$ is only weakly dependent on $M$.

The redshift dependence at a given halo mass, 
using $V^3 \propto \Mv (1+z)^{3/2}$ and $\dot{M} \propto (1+z)^2$, 
is 
\be
\epsfftwo \propto s^{-3/4} (1+z)^{1/2} \, .
\ee
The system can adjust the SFR to match the rate of gas supply by accretion
with $\epsfftwo\sim 1$ at all times
by slight variations in the contraction factor $s$.
At higher redshift, the clumps should contract a bit further and form stars 
at a somewhat higher surface density.

\subsection{Sub-Resolution Clumping}
\label{sec:subclumps}


Our discussion of clump survival in the preceding sections is based on the 
assumption that giant clumps represent single star-forming molecular 
clouds, although we of course expect them to possess significant 
substructure, as do local molecular clouds. 
These substructures are unresolved by current observations and by simulations.
Here we discuss how their presence affects our conclusions.

First note that, for this purpose, we do not care whether
any sub-clumps within the giant clumps themselves survive 
star formation feedback and form bound stellar clusters.
To see why, consider an extreme case in 
which all the sub-clumps
within the giant clump expel most of  their gas, and thus do not leave
behind bound remnants. This is what we might expect to happen if
all the sub-clumps had surface densities similar to that of their parent
giant clump, but had masses well below the $\sim 10^5-10^6$ 
$\msun$ minimum survival mass that we computed in
\S~\ref{sec:survival}. In this case the sub-clumps would all form
stars, expel their gas, and disperse, but both the stars and the
expelled gas would still remain trapped within the much larger
gravitational potential well of the giant clump. They could escape
from this potential well only if the giant clump as a whole were
disrupted by gas expulsion, which we have already shown in
\S~\ref{sec:survival} will happen only if $\epsff$ is much larger
than the expected value. Thus the end result
of this scenario would be a
bound giant star cluster without any bound sub-clusters inside it.

At the opposite extreme, suppose that all the sub-clumps were to remain
bound and undergo negligible gas expulsion. We might expect this
scenario if the sub-clumps all had surface densities much higher
than that of their parent giant clump. In this case the sub-clumps
would convert most of their mass to stars, forming bound clusters.
All the bound clusters would irradiate the remaining mass
in the giant clump, imparting momentum to it. If the stars imparted
enough momentum, this gas would be expelled. Assuming most of the
mass were in the inter-clump medium, as is the case for local
molecular clouds, this expulsion would unbind the giant clump,
producing many small individually bound clusters that are not bound
to one another. Conversely, if the imparted momentum
were not sufficient to unbind the giant clump, as we expect,
the result would be a giant star cluster consisting of many smaller
bound clusters, all gravitationally bound to one another.

In either extreme scenario, whether or not sub-clumps survive does not
make any difference to whether a giant clump as a whole
survives. This is dictated solely by the expulsion fraction from
the giant clump. However, sub-clumping still could make a difference
for giant clump survival by raising the value of $\epsff$. In this case the
sub-clumps would still have $\epsfftwo\sim 1$, but the giant clump
would have $\epsfftwo\gg 1$ because it would have the same star
formation rate but a much lower mean density, and thus a longer
free-fall time.

To see whether this is likely to happen, we note that the
turbulent motions within a giant clump are likely to break it
up into smaller sub-clumps, much a local molecular clouds are broken
up into clump, filamentary structures by turbulence.
In such a configuration, a majority
of the mass is at a density higher than the volumetric mean density $\bar{\rho}$
that we have computed, and would therefore have a shorter free-fall
time and a higher star formation rate. Quantitatively, we have computed
the star formation rate as $\dot{M}_* = \epsff M/t_{\rm ff}(\bar{\rho})$, where
$M$ is the total mass of the giant clump, $\bar{\rho}$ is its
volume-averaged density, and $t_{\rm ff}$ is the free-fall time computed at
that density.
However, if most of the mass is at a density $\rho > \overline{\rho}$,
the appropriate
mass might be the mass $M(>\rho)$ above that higher density,
and the appropriate
timescale might be $t_{\rm ff}(\rho)$ computed for that density.
While one might worry that this could be a significant effect,
\citet{krumholz07g} point out that it is in reality quite small.
Turbulent systems generally have
lognormal density distributions. For such a distribution, the fraction
of the cloud mass with density greater than $\rho$ is given by
\begin{equation}
\frac{M(>\rho)}{M} = \frac{1}{2}\left[1 + 
\mbox{erf}\left(\frac{-2\ln x+\sigma_{\rho}^2}{2^{3/2}\sigma_{\rho}}\right)\right],
\end{equation}
where $x = \rho/\bar{\rho}$, $\bar{\rho}$ is the volumetric
mean density, and $\sigma_{\rho}$ is the dispersion of the density
distribution. This is related to the Mach number $\mathcal{M}$ of
the turbulence by $\sigma_{\rho}^2 \approx \ln (1 + \gamma \mathcal{M}^2)$,
where $\gamma$ is a constant of order unity \citep{padoan02, federrath08a}.
For $\sigma_{\rho} = 2.5 - 3$, 
the range of values expected for the Mach numbers
found in giant clumps, the quantity $M(>\rho)/t_{\rm ff}(\rho)$ varies by only
a factor of a few over a range of densities
$\rho/\bar{\rho} \approx 10^{-1} - 10^5$. Thus even if most of the mass is
at a density vastly larger than the mean density we have used, as long as
the mass distribution follows the lognormal form expected for supersonic
turbulence, the star formation rate will not be modified significantly from our
estimate using the volume-averaged density.

\section{Discussion}
\label{sec:discussion}

\subsection{Turbulence and Energy Balance in Giant Clumps}

Our finding that the fraction of gas ejected from giant clumps depends 
critically on their dimensionless 
star-formation rate efficiency
$\epsff$ naturally leads to the question of how this 
quantity is set, and whether the physical processes responsible for 
setting $\epsfftwo \sim 1$ in the local universe might determine a 
different value in high-redshift clumps. \citet{krumholz05c} show that, 
as long as the gas in 
a molecular cloud is supersonically turbulent with a velocity dispersion 
comparable to the cloud's virial velocity, as is observed to be the case 
in all molecular clouds in the local universe,  
$\epsfftwo \sim 1$ is the inevitable consequence. 
In contrast, in the absence of supersonic turbulence, simulations find 
that clouds undergo a rapid global collapse in which they convert all 
their mass into stars in roughly a dynamical time, i.e. 
$\epsfftwo \sim 100$ \citep[e.g.][]{nakamura07, wang09a}.\footnote{ 
This can be avoided if clouds are magnetically subcritical 
\citep[e.g.][]{nakamura08a}, but magnetic fields in the early universe are 
likely to be weaker than those in the local universe, and even in the local 
universe clouds do not appear to be subcritical in typical clouds 
\citep{crutcher09a}.}
Thus, a value of $\epsfftwo \sim 1$ may be expected in high-$z$ giant 
clumps only if they maintain the level of 
turbulence required to avoid rapid, global collapse.

Whether the turbulence can actually be maintained is somewhat less clear.
Simulations show that supersonic turbulence decays in roughly one
cloud-crossing time \citep[e.g.][]{stone98, maclow98, maclow99}, so global
collapse can be avoided only if this energy is replaced on a comparable 
timescale. In local, low-mass star-forming clouds
($\la 10^4$ $\msun$), observations \citep{quillen05}, simulations 
\citep{nakamura07, wang09a}, and analytic theory \citep{matzner07} 
all suggest that protostellar outflows can supply the necessary energy. 
In local giant molecular clouds with masses of 
$\sim 10^4 - 10^6$ $\msun$, \hii\ regions driven by the pressure of 
photoionized gas are likely to be able to supply the necessary energy 
\citep{matzner02, krumholz06d}. However, neither of these mechanisms are 
effective for clumps with $M_9\sim 1$ and $\Sigma_{-1}\sim 1$, because they 
do not provide enough momentum input and because they are overwhelmed by 
radiation pressure (see Figure 2 of \citealt{fall10a}). 

Supernova feedback \citep{dekel86a} 
does not appear to be a likely candidate to drive the
turbulence either. Supernovae do not provide enough power
to drive the observed level of turbulence \citep{dekel09b},
and analytic calculations \citep{harper-clark09a, krumholz09d},
numerical simulations of isolated disk galaxies 
\citep{tasker08a, joung09a}, and numerical simulations of galaxies
in cosmological context \citep{ceverino09a} all indicate that
supernova-heated gas is likely to escape through low-density 
holes in the molecular gas without driving much turbulence. 

Contrary to this conclusion,
\citet{lehnert09a} use the observed correlation between
H$\alpha$ surface brightness and linewidth in $z\sim 2$ galaxies
to argue that supernova feedback is responsible for driving
the turbulence, based in part on simulations by \citet{dib06a}, who
obtain a scaling relation between velocity dispersion and supernova
rate in numerical simulations. However, the efficiency with which
supernova energy is coupled to the ISM is a free parameter
in both \citeauthor{lehnert09a}'s analysis and in \citeauthor{dib06a}'s
simulations, and their results are consistent with the data only if it
is $\sim 25\%$, whereas in the dense environments found in
high redshift galaxies it is expected to be far lower \citep{thompson05}.
This conclusion is confirmed by the more recent simulations, which
do not need to assume an efficiency 
because they have sufficient resolution to resolve the multiphase
structure of the ISM. Finally, we note that the
correlation between H$\alpha$ surface brightness and linewidth 
observed by \citeauthor{lehnert09a} has a more prosaic explanation:
in a marginally stable galactic disk of constant circular velocity,
the velocity dispersion is proportional to the gas surface density, 
since $Q=\kappa\sigma/(\pi G \Sigma) = 1$. Thus higher velocity
dispersions correspond to higher surface densities, which in turn
produce higher star formation rates in accordance with the standard
\citet{kennicutt98a} relation. This naturally explains the observed
correlation.

Radiation pressure is another mechanism to consider.
If radiation pressure is not able to drive mass out of the clump, 
as found above for $\epsff \sim 0.01$,
this suggests that it might not be able to drive turbulence to the required 
virial level either, since the virial and escape velocities only differ by 
a factor of $\sqrt{2}$. However, it is unclear whether
this conclusion is warranted. Radiation pressure cannot drive material out 
of a clump not because stars do not accelerate material enough, but because 
they evolve off the main sequence before they are actually able to eject 
matter. As a result, they produce expanding shells whose velocities greatly 
exceed the escape velocity. They simply fail to drive mass out because the 
clump because the driving sources turn off before the shells actually 
escape from the cluster. It is unclear if the expanding shells might 
provide enough energy to maintain the turbulence; this problem will 
require further modeling.

We are left with the possibility that the turbulence is driven by gravity.
The driving source {\it cannot} be the collapse of
the clump itself; although such a collapse does produce turbulence, 
it does so at the price of reducing the crossing time, raising the rate of 
energy loss. Consequently, the collapse becomes a runaway process, and all 
the gas quickly converts to stars. However, as shown by \citet{dekel09b}, 
the gravitational migration of the clumps {\it through} the galactic disc 
does provide enough power to maintain the turbulence within them. 
The main uncertainty in this model is how much of that power will go into 
driving internal motions within the clump, rather than motions in the 
external galactic disc. This depends on how the clumps are torqued by 
one another and by the disc. However, there is suggestive evidence from 
simulations of giant clumps that this mechanism might be viable. 
The simulations of clumpy galaxies that have been done to date 
\citep[e.g.][]{bournaud07a, elmegreen08b, agertz09a, ceverino09b}  
either include no feedback or only supernova feedback (which is 
ineffective). 
If the giant clumps formed in these simulations did lose their turbulence 
and undergo global collapse, then, depending on the details of the
simulation method, they would either convert all of their mass into 
stars, or all of the mass within them would collapse to the maximum 
density allowed by the imposed numerical pressure floor. 
This collapse would happen on the crossing 
time scale of a clump, which is much less than the time required for the 
clumps to migrate to the galactic centre. However, such collapses are not 
observed in the simulations. This strongly suggests that, even in the 
absence of feedback, gravitational power is sufficient to maintain the 
turbulence.

We conclude that the generation of turbulence in the giant clumps is 
an important open issue, to be addressed by further studies including
simulations of higher resolution.

\subsection{Comparison to Previous Work}
\label{murraymass}

Our conclusion 
that giant clumps are not likely to be disrupted by feedback is 
in contrast with the findings of \citet{murray09a} 
for the giant clump in the galaxy Q2346-BX 482 \citep{genzel08a}, 
and this difference merits discussion. Based on the H$\alpha$ luminosity of 
the clump, \citeauthor{murray09a} estimate a total bolometric luminosity of 
$L=4\times 10^{11}$ $\lsun$, which corresponds to a star-formation rate of 
34 $\msun$ yr$^{-1}$ in the old stars limit, or a stellar mass of 
$2.6\times 10^8$ $\msun$ in the young stars limit, 
which \citeauthor{murray09a} assume. If these stars are indeed young, 
then the star-formation rate could be higher than 34 $\msun$ yr$^{-1}$, 
but it could not be any lower.

While the estimate of the star-formation rate is relatively 
straightforward, inferring the gas mass is much less so.  
\citet{murray09a} take it to be 
$10^9$ $\msun$ based on an order-of-magnitude estimate for the Toomre 
mass in the galaxy. This choice is crucial to their result. 
The clump radius is $r=925$ pc, 
so if we adopt this radius, the mean density-free fall time is 15 Myr, 
so $\epsfftwo \simeq 50$. 
Using this value in equation (\ref{eff2}), together with the corresponding 
surface density and mass 
$\Sigma_{-1} = 0.8$ and $M_9 = 1$, 
tells us that we are in the young stars limit and that 
$\calE = 0.06$
 -- fully consistent with \citeauthor{murray09a}'s conclusion that only a 
relatively small fraction of the gas mass turns into stars, and that this 
is sufficient to expel the remaining gas. \citeauthor{murray09a} derive a 
slightly higher value $\calE \sim 1/3$ because their criterion for ejection 
amounts to adopting $\alpha_{\rm crit}\sim 10$.

However, this conclusion depends crucially on having a low estimate of 
the gas mass, and a correspondingly high estimate for $\epsfftwo$. 
Indeed, if the value of $\epsfftwo \simeq 50$ were accurate, this clump would 
have the highest star formation rate efficiency of any known system.
It is therefore useful to consider alternative methods for estimating the mass.
If we were to adopt the inverse-Kennicutt method, the observed 
star-formation rate per unit area
$\dot{\Sigma}_* \simeq 13$ $\msun$ yr$^{-1}$ kpc$^{-1}$, 
together with equation (\ref{kennicutt}), gives a gas surface density of 
$\Sigma_{-1} \simeq 5$ (or 2300 $\msun$ pc$^{-2}$). The corresponding gas mass 
is $6\times 10^9$ $\msun$, and recomputing $\epsff$
for this mass gives $\epsfftwo \simeq 3$, 
consistent with the point we made earlier that the Kennicutt Law is in 
practice equivalent to having $\epsfftwo \sim 1$. With this mass we 
would predict $\calE \simeq 0.9$, i.e.\ essentially no gas expulsion. 
One would expect similar results from 
\citeauthor{murray09a}'s models, because the effect of this mass increase 
would be to increase the gravitational force by a factor of 40 while 
leaving the radiative force unchanged. Thus while \citeauthor{murray09a}'s 
models suggest that the clump in BX 482 has stopped forming stars and all 
the remaining gas has just been expelled, according to our estimate
this
clump can be only part of the way through its life and may continue 
to form stars.

\citet{murray09a} also suggest another method of estimating the gas mass. 
Based on the H$\alpha$ luminosity, if one assumes that the clump is filled 
with uniform-density gas, then ionization balance requires that this 
gas have a density of hydrogen nuclei
\begin{equation}
n_{\rm H} = \sqrt{\frac{3 S}{4\pi r^3 \alpha^{(B)}}\left(\frac{4X}{3X+1}\right)},
\end{equation}
where $X$ is the hydrogen mass fraction and this expression assumes that 
He is singly-ionized.
If, following \citet{murray09a}, we adopt the young stars limit, then the 
ionizing luminosity is 
$S=6.3\times 10^{46} (M_*/\msun) = 1.6\times 10^{55}$ 
photons s$^{-1}$ \citep{murray09c}.\footnote{The ionizing luminosity is 
somewhat lower in the old stars limit: $S = 2.5\times 10^{53} 
(\dot{M}_*/\msun\mbox{ yr}^{-1}) = 8.5\times 10^{54}$ photons s$^{-1}$.} 
Combining this with the case B recombination coefficient 
$\alpha^{(B)}=3.46\times 10^{-13}$ cm$^3$ s$^{-1}$ and the Solar hydrogen 
mass fraction $X=0.71$ (mean mass per H nucleus of $1.4m_{\rm H}$) gives 
$n_{\rm H} = 21$ cm$^{-3}$, 
corresponding to a mass of $2.4\times 10^9$ $\msun$
and $\epsfftwo = 14$. Plugging this mass, surface density, and value of 
$\epsfftwo$ into equation (\ref{eff2}) gives $\calE \simeq 0.2$, i.e.\ the  
star fraction is more than three times what we would obtain using 
\citeauthor{murray09a}'s mass of $M=1\times 10^9$ $\msun$. 
Adopting $\alpha_{\rm crit}=10$ to shift 
our fiducial parameters closer to those used in \citeauthor{murray09a} 
would give $\calE \simeq 0.8$, no significant gas expulsion. We emphasize that 
these calculations are lower limits on the gas mass and upper limits on the 
fraction of mass ejected, because this mass estimate includes only 
ionized gas.  However, models of both classical gas pressure-driven 
\hii\ regions and ones driven by radiation pressure 
\citep{krumholz09d, murray09a} suggest that the 
ionized gas mass in the \hii\ region interior is significantly smaller 
than the mass of neutral gas swept up in the shell around it. 
Including this mass would lower $\epsfftwo$ and increase $\calE$ 
even further.

In summary, our conclusions differ from those of \citet{murray09a} 
not because of any difference in the physics of radiation feedback, 
but because they have used an estimated gas mass that produces an 
extraordinarily high value of $\epsff$.

\section{Conclusions}
\label{sec:summary}

Our main result in this paper, summarized in Equation (\ref{eq:expulsion})
and Figure \ref{fig:starfrac},
is that the survival or disruption of giant star-forming clumps in
high-$z$ galaxies depends critically on the rate at which they turn
into stars.
We find that as long as the high-redshift clumps convert their gas mass
into stars at a rate of one to a few percent of the mass per free-fall
time, $\epsff \sim 0.01$,
as is observed in low-redshift star-forming systems from small galactic
clusters to ultraluminous infrared galaxies,
the clumps retain most of their gas and turn it into stars.
As a result, they remain bound as they migrate into the
galactic centre on timescales of $\sim 2$ disc orbital times.
A significant fraction of the clump gas could be ejected on a free-fall
timescale before turning into stars only if
clumps can convert $\ga 10\%$ of their gas mass into stars in a
free-fall time, forming stars much faster than any other
star-forming system known.
We argue the current high-$z$ data is consistent with the standard SFR rate
efficiencies at the level of one to a few percent, with
no significant evidence for a change in the star formation process in
high-$z$ star-forming galaxies. Nevertheless, this is clearly an interesting
issue to explore with more direct observational estimates of the gas
mass in these galaxies.

It is possible to check our theoretical arguments for clump survival
with a number of possible observations. First, clumps can be
disrupted by radiation pressure only if a majority of their gas is expelled,
and the resulting
massive radiation-driven outflows from
clumps may be observable as systematic blueshifts at
the clump locations. A preliminary search for such a phenomenon have
yielded a null result (K. Shapiro, private communication, 2009), but further
investigations of the kinematics in and around the clumps are worthwhile.
It is possible that sufficient extinction could hide the blueshifted
signature, but we note that, if the gas surface densities are relatively
low as, e.g., \citet{murray09a} propose, the extinction is relatively mild.

Second, stellar populations in high-redshift clumps can
provide independent observational tests that could help distinguish between
the two scenarios of clump survival or disruption.
One such test involves the age spread of stars in actively 
star-forming clumps. If radiative feedback disrupts the clumps after one or 
a few free-fall times, the spread of stellar ages in each clump should
not exceed $\sim 50$ Myr. If clumps survive, on the other hand, then the age 
spreads may reach $\ga 100$ Myr,
with a high and roughly constant SFR during the lifetime of the clump.
There is preliminary observational evidence in favor of the latter
\citep{elmegreen09c, forster-schreiber09a}, but this should be explored further.
In this case, migration toward the bulge 
will produce an age gradient, so that clumps closer
to the bulge have systematically larger age spreads than those
further out in the disc.

A third test is associated with the properties of
massive star clusters in the disc that are not actively forming stars.
Clusters with masses $\ga 10^7$ $\msun$ can only be made
in giant clumps, not in the rest of the disc, where molecular
clouds are smaller. If clumps undergo rapid disruption by
feedback, most of the clusters are likely to dissolve as is
the case for local star formation \citep[e.g.][]{fall09a}, but
the few clusters that may survive will remain in the disc near
their formation radii.
None will migrate into the bulge, since the migration
time varies as $M^2$ \citep{dekel09a}, and the masses of the
clusters are much smaller than those of their parent gas clumps.
In contrast, if gas clumps survive feedback and convert most of
their mass to stars, the resulting massive objects will rapidly 
migrate toward the galactic center. They will lose $\sim 50\%$
of their mass due to tidal stripping \citep{bournaud07a}, possibly
including some massive sub-clusters, but they will deliver the
rest of their mass to the bulge. Thus, if giant clumps do not
survive, we expect to see a few massive clusters in the disk
and none in the bulge. If, on the other hand, our model is
correct, then there should be comparable masses of disk
and bulge clusters. Those bulge clusters that survive today
may correspond to the metal-rich globular cluster
population seen in present-day galactic bulges
(\citealt{brodie06a, shapiro10a}; Romanowsky et al., in preparation).


Even though we have concluded that the high-$z$ giant clumps are expected to
survive radiative stellar feedback, we do emphasize the important role 
that this process is likely to play in galaxy formation. For example, 
it can provide some of the pressure support needed in these giant clumps, 
it can drive non-negligible winds out of them, and it is likely to disrupt 
the less massive clumps where stars form at later 
redshifts.  In many circumstances, this mode of feedback is expected to be more
important than supernova feedback, as it pushes away the cold dense gas
while the latter mostly affects the hot dilute gas. We thus highlight the need
for incoroprating radiative stellar feedback in hydrodynamical simulations
as well as in semi-analytic models of galaxy evolution.


\section*{Acknowledgments}

We acknowledge stimulating discussions with Andi Burkert, Daniel
Ceverino, Reinhard Genzel, Norm Murray, and Aaron Romanowsky,
and helpful comments on the manuscript from Mark Dijkstra,
Bruce Elmegreen, and the referee, Frederic Bournaud.
This research has been supported by the Alfred P.\ Sloan Foundation (MRK); 
NASA through ATP grants NNX09AK31G (MRK), NAG5-8218 (AD), 
and as part of the Spitzer Theoretical Research Program, through a contract issued by the JPL (MRK); 
the National Science Foundation through grant AST-0807739 (MRK); 
the German-Israeli Foundation (GIF) through grant I-895-207.7/2005 (AD);
the German Research Foundation (DFG) via German-Israeli Project Cooperation
grant STE1869/1-1.GE625/15-1 (AD); the Israel Science Foundation (AD); 
and France-Israel Teamwork in Sciences (AD).

\label{lastpage}

\bibliographystyle{mn}
\bibliography{refs}

\end{document}